\documentclass[preprint,review,12pt]{elsarticle}
\biboptions{sort&compress}
\usepackage{graphics}
\usepackage{graphicx}
\usepackage{amssymb}
\usepackage{color}
\usepackage{graphicx}
\usepackage{epstopdf}
\usepackage{amsmath}
\usepackage{appendix}

\journal{CARBON}

\begin{document}

\begin{frontmatter}
\title{Superconductivity in Ca-intercalated bilayer graphene: C$_{2}$CaC$_{2}$}
\author{Jin-Han Tan$^a$, Hao Wang$^a$, Ying-Jie Chen$^a$, Na Jiao$^a$}
\author{ Meng-Meng Zheng$^{a,*}$}
\ead{qfzhmm@163.com}
\author{ Hong-Yan Lu$^{a,*}$}
\ead{hylu@qfnu.edu.cn}
\author{ Ping Zhang$^{a,b,**}$}
\ead{pzhang2012@qq.com}
\address{$^a$School of Physics and Physical Engineering, Qufu Normal University, Qufu 273165, China \\$^b$Institute of Applied Physics and Computational Mathematics, Beijing 100088, China}

\begin{abstract}
\indent The deposition and intercalation of metal atoms can induce superconductivity in monolayer and bilayer graphenes. For example, it has been experimentally proved that Li-deposited graphene is a superconductor with critical temperature $T_{c}$ of 5.9 K, Ca-intercalated bilayer graphene C$_{6}$CaC$_{6}$ and K-intercalated epitaxial bilayer graphene C$_{8}$KC$_{8}$ are superconductors with $T_{c}$ of 2-4 K and 3.6 K, respectively. However, the $T_{c}$ of them are relatively low. To obtain higher $T_{c}$ in graphene-based superconductors, here we predict a new Ca-intercalated bilayer graphene C$_{2}$CaC$_{2}$, which shows higher Ca concentration than the C$_{6}$CaC$_{6}$. It is proved to be thermodynamically and dynamically stable. The electronic structure, electron-phonon coupling (EPC) and superconductivity of C$_{2}$CaC$_{2}$ are investigated based on first-principles calculations. The EPC of C$_{2}$CaC$_{2}$ mainly comes from the coupling between the electrons of C-$p_{z}$ orbital and the high- and low-frequency vibration modes of C atoms. The calculated EPC constant $\lambda$ of C$_{2}$CaC$_{2}$ is 0.75, and the superconducting $T_{c}$ is 18.9 K, which is much higher than other metal-intercalated bilayer graphenes. By further applying -4\% biaxial compressive strain to C$_{2}$CaC$_{2}$, the $T_{c}$ can be boosted to 26.6 K. Thus, the predicted C$_{2}$CaC$_{2}$ provides a new platform for realizing superconductivity with the highest $T_{c}$ in bilayer graphenes.\\
\end{abstract}
   \end{frontmatter}

\section{Introduction}
\indent	It is well-known that graphene is a two-dimensional (2D) Dirac semimetal characterized by a complete absence of electronic density of states (DOS) at the Fermi energy. Since its isolation in 2004 \cite{ref:1}, graphene has attracted significant attention from both experimental and theoretical researchers. This has resulted in the exploration of numerous intriguing properties that are not typically observed in ordinary 2D electron gases. Although graphene exhibits semi-metallic characteristics, its properties can be modified to transform into a metallic state through appropriate modulations. The realization of superconductivity in graphene has been the subject of extensive research. For instance, in previous years, the topic of intense discussion has been the magic angle bilayer graphene, which exhibits the $T_{c}$ of 1.7 K with twist angles of 1.1° for bilayer graphene \cite{ref:2,ref:3}. Few-layer graphene or graphite stacked from polylaminate graphene can also be regulated to possess metallicity. Actually, superconductivity has been extensively studied and experimentally confirmed in graphite intercalation compounds (GICs), which can be synthesized by introducing metal dopants into bulk graphite \cite{ref:4,ref:5,ref:6}. Representative GICs include CaC$_{6}$ \cite{ref:7} and YbC$_{6}$ \cite{ref:8}, which exhibit $T_{c}$ of 11.5 and 6.5 K, respectively. For nearly two decades, researchers have been committed to the search for related graphene based superconductors, driven by their irreplaceable superior physical properties and extensive potential applications. If a high superconducting $T_{c}$ can be achieved in graphene based superconductors, it would significantly contribute to the development of nano superconducting devices.\\
\indent	Graphene exhibits robust 2D properties and possesses a multitude of remarkable characteristics, including the quantum Hall effect \cite{ref:9}, exceptional mechanical strength and flexibility \cite{ref:10}, and so on \cite{ref:11,ref:12,ref:13,ref:14,ref:15,ref:16,ref:17,ref:18,ref:19,ref:20}. For the fabrication of 2D superconducting graphene, the process involves the deposition of alkali metal or alkaline-earth metal onto monolayer graphene, as well as the intercalation of alkali metal atoms into bilayer graphene \cite{ref:21}. For example, alkali metal Li-deposited graphene LiC$_{6}$ and alkaline-earth metal Ca-deposited graphene CaC$_{6}$ were predicted to be phonon-mediated superconductors with superconducting $T_{c}$ of 8.1 and 1.4 K \cite{ref:22}, respectively. Moreover, after applying 10\% biaxial tensile strain, it was predicted that the $T_{c}$ of LiC$_{6}$ increases from 8.1 K to about 28.7 K \cite{ref:23}. Subsequent experiments have confirmed the superconductivity of LiC$_{6}$ at a temperature of 5.9 K \cite{ref:24}. It has been predicted that metals-deposited graphene, such as AlC$_{8}$,  could exhibit superconductivity with $T_{c}$ exceeding 22 K at the experimentally accessible hole doping and tensile strain levels \cite{ref:25}. The $T_{c}$ of the other successfully synthesized compound KC$_{8}$ is 0.55 K \cite{ref:26}. In addition to the metal-deposited graphene superconductors, significant advancements have also been made in the field of metal-intercalated bilayer graphene superconductors. Ca-intercalated bilayer graphene C$_{6}$CaC$_{6}$ \cite{ref:27} was experimentally prepared with a $T_{c}$ of 4 K. Subsequent theoretical calculations revealed that C$_{6}$CaC$_{6}$ possesses the capability to sustain phonon-mediated superconductivity, with $T_{c}$ ranging from 6.8 to 8.1 K. This theoretical prediction is in excellent agreement with the experimental data \cite{ref:28}. Recently, an experimental study successfully prepared the K-intercalated bilayer graphene, and the resulting $T_{c}$ was determined to be 3.6 K \cite{ref:21}. However, the $T_{c}$ of them are relatively low. Whether the $T_{c}$ of bilayer graphene superconductors can be increased is an open question.\\
\indent	In this work, we predict a new calcium intercalated bilayer graphene compound, namely Ca intercalated AA-stacking bilayer graphene C$_{2}$CaC$_{2}$, which shows higher Ca concentration than C$_{6}$CaC$_{6}$. This prediction is grounded in the experimental synthesis of previously known compounds. To validate the precision of our calculation, we performed calculations on the 2D calcium intercalated bilayer graphene C$_{6}$CaC$_{6}$ with varying intercalation concentrations. The calculated $T_{c}$ was approximately 8 K, which closely aligned with both theoretical predictions \cite{ref:26} and experimental \cite{ref:25} findings. In this paper, the electronic structure, EPC and superconductivity of C$_{2}$CaC$_{2}$ are investigated based on first-principles calculations. Based on the BCS theory, it has been determined that the EPC constant $\lambda$ of C$_{2}$CaC$_{2}$ is 0.75, and the superconducting $T_{c}$ is 18.9 K, which is higher than other metal-intercalated bilayer graphenes. The EPC of C$_{2}$CaC$_{2}$ primarily comes from the coupling between the electrons of C-$p_{z}$ orbital and the high- and the low-frequency vibration modes of C atoms. Then, we further apply -4\% biaxial compressive strain to C$_{2}$CaC$_{2}$, resulting in an increase of the $T_{c}$ to 26.6 K. Our prediction of C$_{2}$CaC$_{2}$ presents a novel avenue for the exploration of bilayer graphene superconductors with the highest $T_{c}$.\\
\indent	The rest of this paper is organized as follows. In Section II, the computational details are described. In Section III, the results and discussions are presented. The crystal structure and stability of C$_{2}$CaC$_{2}$ are presented, with stability being characterized in terms of both thermodynamic and dynamical stability. Secondly, the electronic structure of C$_{2}$CaC$_{2}$ is studied, including the band structure, density of states (DOS), and bader charge of C$_{2}$CaC$_{2}$. Thirdly, the phonon properties and EPC of pristine and biaxial strained C$_{2}$CaC$_{2}$ are calculated and the possible superconducting $T_{c}$ is then calculated. Section IV is the conclusion of the study.\\

\section{Computational details}
\indent	All calculations related to C$_{2}$CaC$_{2}$ in this article are performed in the framework of density functional theory (DFT), as implemented in the Vienna ab initio simulation package (VASP) \cite{ref:29} and the Quantum Espresso (QE) program \cite{ref:30}. The interaction between electrons and ions is realized by the Projector-Augmented-Wave (PAW) method \cite{ref:31}, and the exchange-correlation potentials are treated using the generalized gradient approximation (GGA) with Perdew-Burke-Ernzerhof (PBE) parametrization \cite{ref:32}. Both the lattice parameters and the atom positions are relaxed to obtain the optimized structure. The energy cutoffs for wave functions and charge density are set as 80 Ry and 800 Ry, respectively. Electronic integration is carried out on an 18$\times$18$\times$1 $k$-point grid. For the calculation of the DOS, $k$ points of 36$\times$36$\times$1 are used. The phonon and EPC are calculated on a 12$\times$12$\times$1 $q$-point grid, and a denser 48$\times$48$\times$1 $k$-point grid is used for evaluating an accurate electron-phonon interaction matrix.\\
\indent The total EPC constant $\lambda$ is obtained via the isotropic Eliashberg function \cite{ref:33,ref:34,ref:35}:
	\begin{eqnarray}
	\alpha^{2}F(\omega)=\frac{1}{2\pi N(E_{F})}\sum_{\mathbf{q}\nu}\delta(\omega-\omega_{\mathbf{q}\nu})\frac{\gamma_{\mathbf{q}\nu}}{\omega_{\mathbf{q}\nu}},
\end{eqnarray}
\begin{eqnarray}
	\lambda=2\int_{0}^{\infty}\frac{\alpha^{2}F(\omega)}{\omega}\,d\omega=\sum_{\mathbf{q}\nu}^{}\lambda_{\mathbf{q}\nu},
\end{eqnarray}
where $\alpha^{2}$$F$($\omega$) is Eliashberg function and $N$($E_{F}$) is the DOS at the Fermi level, $\omega_{\mathbf{q}\nu}$ is the phonon frequency of the $\nu$th phonon mode with wave vector $\mathbf{q}$, and $\gamma_{\mathbf{q}\nu}$ is the phonon linewidth \cite{ref:33,ref:34,ref:35}. The $\gamma_{\mathbf{q}\nu}$ can be estimated by
\begin{eqnarray}
	\begin{split}
		\gamma_{\mathbf{q}\nu}=\frac{2\pi\omega_{\mathbf{q}\nu}}{\Omega_{BZ}}\ \sum_{\mathbf{k},n,m}\lvert g^{\nu}_{\mathbf{k}n,\mathbf{k}+\mathbf{q}m}\rvert^{2}\delta(\epsilon_{\mathbf{k}n}-E_{F})\\\delta(\epsilon_{\mathbf{k}+\mathbf{q}m}-E_{F}),
	\end{split}
\end{eqnarray}
where $\Omega$$_{BZ}$ is the volume of the BZ, $\epsilon_{\mathbf{k}n}$ and $\epsilon_{\mathbf{k}+\mathbf{q}m}$ indicate the Kohn-Sham energy, and $g^{\nu}_{\mathbf{k}n,\mathbf{k}+\mathbf{q}m}$ represents the screened electron-phonon matrix element. $\lambda_{\mathbf{q}\nu}$ is the EPC constant for phonon mode $\mathbf{q}\nu$, which is defined as 
\begin{eqnarray}
	\lambda_{\mathbf{q}\nu}=\frac{\gamma_{\mathbf{q}\nu}}{\pi\hbar N(E_{F})\omega_{\mathbf{q}\nu}^{2}}\,
\end{eqnarray}
$T_{c}$ is estimated by McMillan-Allen-Dynes formula \cite{ref:35}:
\begin{eqnarray}
	T_{c}=\frac{\omega_{log}}{1.2}exp[\frac{-1.04(1+\lambda)}{\lambda-\mu^{*}(1+0.62\lambda)}].
\end{eqnarray}
The hysteretic Coulomb pseudopotential $\mu^{*}$ in Eq.(5) is set to 0.1 and logarithmic average of the phonon frequencies $\omega_{log}$ is defined as
\begin{eqnarray}
	\omega_{log}=exp[\frac{2}{\lambda}\int_{0}^{\omega_{}}\alpha^{2}F(\omega)\frac{log\;\omega}{\omega}d\omega].
\end{eqnarray}
\section{Results and Discussions}
\subsection{Crystal structure and stability}

\begin{figure}
	\includegraphics[width=1\linewidth,height=0.7\linewidth]{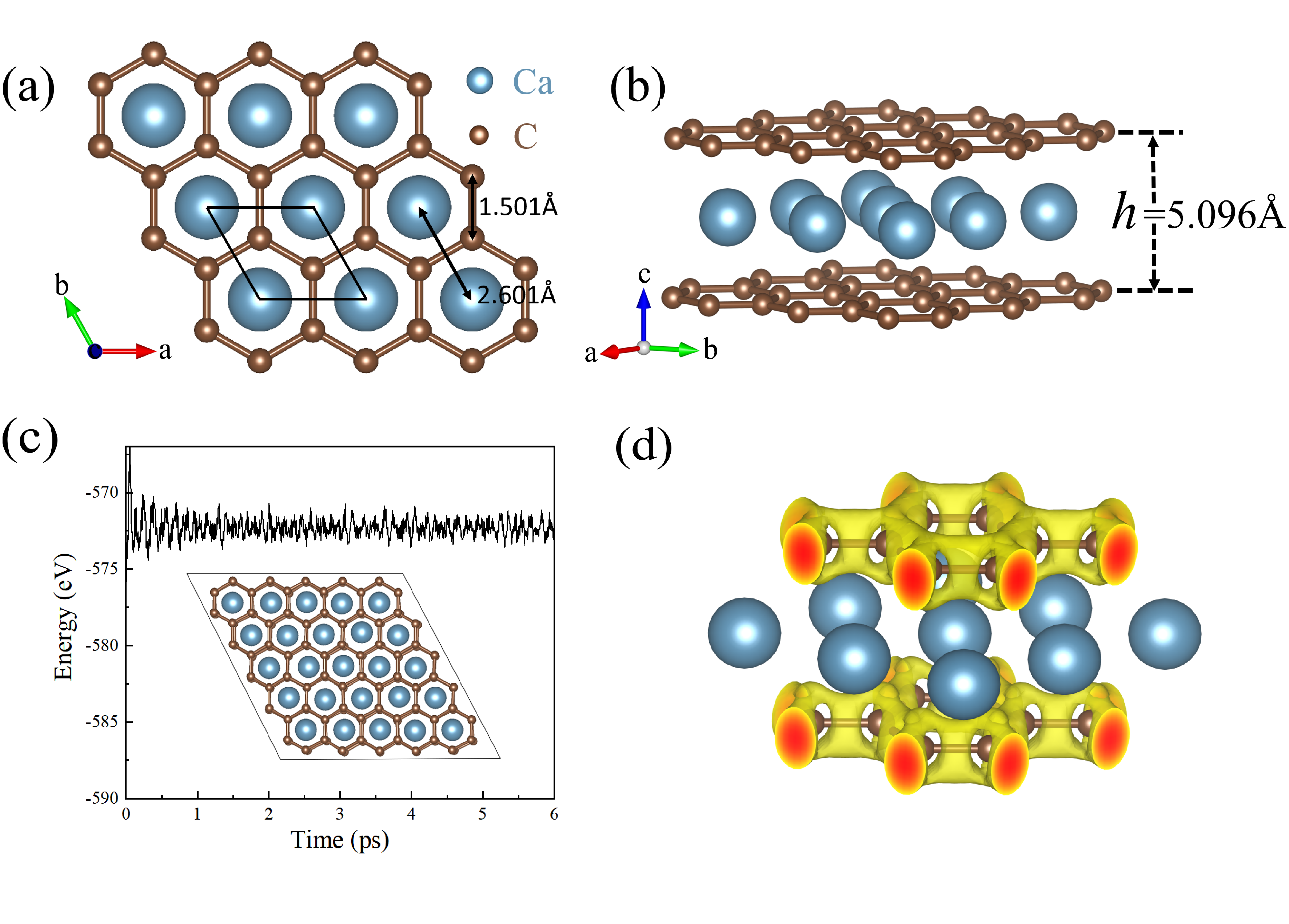}
	\caption{ Top (a) and side (b) views of C$_{2}$CaC$_{2}$. The unit cell is shown by the black solid line. Carbon and calcium atoms are represented by brown and blue spheres, respectively. (c) The variation of the free energy in the AIMD simulations in the time scale of 6 ps along with the last frame of image at 500 K for C$_{2}$CaC$_{2}$. (d) Electron localization function of C$_{2}$CaC$_{2}$.}
	\label{fig:combination1}
\end{figure}

\indent The optimized structure of Ca-intercalated AA-stacking bilayer graphene C$_{2}$CaC$_{2}$ exhibits hexagonal symmetry with the space group of P6/mmm (No. 191). The top and side views of the lattice structure of C$_{2}$CaC$_{2}$ are depicted in Fig. 1(a). The brown and blue spheres correspond to carbon and calcium atoms, respectively. In Fig. 1(a), the intercalated calcium atoms are positioned at the hollow sites of carbon six-membered rings in a honeycomb structure. There are five atoms within each unit cell, including two layers of four carbon atoms, and a calcium atom between the two carbon layers. From the side view in Fig. 1(b), the bilayer is clearly observable. In all calculations, a vacuum space of 15 Å is employed to avoid the interactions between adjacent layers. Graphene, known for its flat structure, maintains its symmetry in the crystal structure even after the intercalation of calcium atoms into the bilayers. The vertical  distance between the bilayers is 5.096 Å as shown in Fig. 1(b). The lattice constant of C$_{2}$CaC$_{2}$ is 2.605 Å. Additionally, the bond lengths of C-C and Ca-Ca are 1.501 Å and 2.601 Å, respectively.\\
\indent The stability of C$_{2}$CaC$_{2}$ is studied from two aspects, i.e., the thermodynamic stability and the dynamical stability. To investigate the thermodynamic stability, it has been demonstrated by the ab initio molecular dynamics (AIMD) simulations. A 4$\times$4$\times$1 supercell is used to minimize the effect of periodic boundary conditions. The variation of free energy during the AIMD simulations within 6 ps, as well as the last frame of the photographs, is exhibited in Fig. 1(c). The results indicate that the structural integrity remains unchanged even at 500 K. Additionally, the energy levels exhibit fluctuations around -573 eV, proving the thermodynamic stability of C$_{2}$CaC$_{2}$. Regarding the dynamic stability of C$_{2}$CaC$_{2}$, the phonon spectrum is calculated, as shown in Fig. 3(a). The absence of imaginary frequency suggests its dynamic stability.\\
\indent Apart from the Ca-intercalated AA-stacking bilayer graphene, we also consider the crystal structure of the AB-stacking case (Fig. A.7), which also contains four carbon atoms and one calcium atom in the unit cell. Although there is no imaginary frequency in the phonon spectrum as shown in Fig. A.8, the AIMD simulation reveals that the AB-stacking C$_{2}$CaC$_{2}$ configuration undergoes a gradual transformation at both low temperature (50 K) and room temperature (300 K), respectively. At 50 K, the structure undergoes a complete phase transition and becomes AA-stacking, the same as that shown in Fig. 1.(a), the AIMD is depicted in Fig. A.9. At 300 K, the bilayer graphene becomes misaligned, and the calcium monolayer turns into two layers. Whether this new structure is stable and its possible properties are beyond the scope of this article. Further research is on going. Therefore, in the following, we only study the properties of AA-stacking C$_{2}$CaC$_{2}$.\\

\subsection{Electronic structure}
\indent We firstly calculate the electronic structure of C$_{2}$CaC$_{2}$ based on first-principles calculations. Fig. 2 shows the orbital-resolved band structures, electronic DOS, and orbital-projected density of states (PDOS). Fig. 2(a) and 2(b) depict the orbital-resolved band structure along the high-symmetry line $\Gamma$–$M$–$K$–$\Gamma$ for C and Ca. Multiple bands cross the Fermi level, indicating the metallic nature of this structure. According to the data presented in Fig. 2(c), the C and Ca atoms exhibit nearly identical contribution to the states around the Fermi level. At the Fermi level, there is a Van Hove singularity peak [Fig. 2(c)], which primarily originating from the saddle point near the $M$ point, predominantly contributed by the C-$p_{z}$ orbital, as can be seen from Fig. 2(a) and 2(d). The Van Hove singularity at the Fermi level is beneficial for superconductivity. According to Fig. 2(d), the primary contribution at the Fermi level is attributed to the C-$p_{z}$ orbitals, with the Ca-$d_{z^{2}}$ and Ca-$d_{xy}/d_{x^{2}-y^{2}}$ orbitals following in significance. And the peak of DOS appears in the same position near the Fermi level, as depicted in Fig. 2(c), this indicated interactions between Ca and C atoms.\\
\begin{figure*}
	\centering
	\includegraphics[width=1.0\linewidth,height=0.5\linewidth]{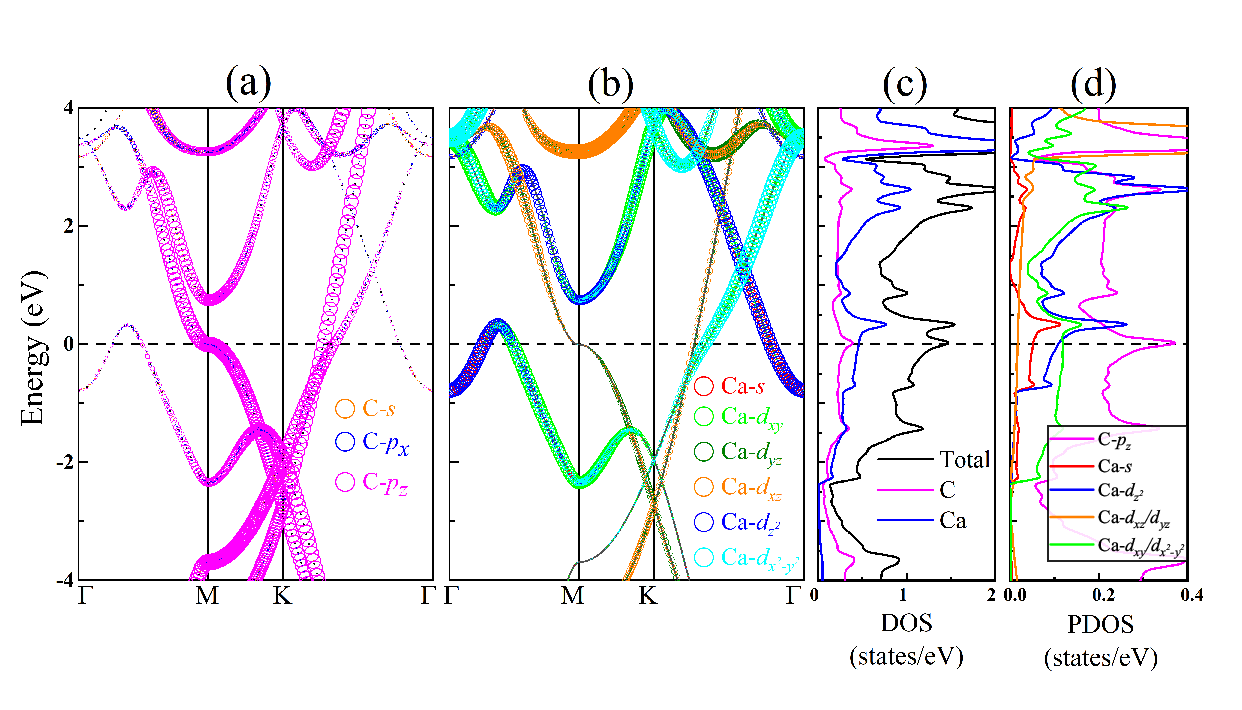}
	\caption{ Orbital-resolved electronic band structure along high-symmetry line of $\Gamma$$-$$M$$-$$K$$-$$\Gamma$ for C (a) and Ca (b) of C$_{2}$CaC$_{2}$. (c) The total DOS of C$_{2}$CaC$_{2}$ and the total DOS of C and Ca atoms. (d) Orbital-projected DOS of C$_{2}$CaC$_{2}$.}
\end{figure*}
\indent The electron-transfer properties of C$_{2}$CaC$_{2}$ are estimated via calculating the Bader charge and electron localization function (ELF). The net charges of the C and Ca atoms were found to be 4.16 and 9.36, respectively. This analysis revealed that each Ca atom transfers 0.64 electrons to the bilayer graphene. The ELF analysis of C$_{2}$CaC$_{2}$ in Fig. 1(d) reveals that the electrons transferred from Ca are predominantly localized in the graphene layers around the C-C bonds. The above results suggest the presence of an ionic bond between the C and Ca atoms, and it is evident that the carbon atoms within the graphene layer are connected by robust covalent bonds.\\

\subsection{Electron-phonon coupling and possible superconductivity}

\indent To investigate the EPC and the potential for superconductivity in C$_{2}$CaC$_{2}$, an analysis of the phonon dispersion, phonon density of states (PhDOS), Eliashberg spectral function $\alpha$$^{2}$$F$($\omega$) and the accumulative EPC constant over the entire frequency spectrum was conducted, which are shown in Fig. 3. As depicted in Fig. 3(a), it can be seen that the contribution of the in-plane and out-of-plane vibration modes of carbon and calcium atoms to the phonon spectrum varies in different frequency ranges. The acoustic branches below 352 cm$^{-1}$ are primarily attributed to the in-plane and out-of-plane vibration modes of Ca. Besides, the out-of-plane vibration modes of C also make a significant contribution to the lowest acoustic branch. At the frequency range of 353 to 564 cm$^{-1}$, the out-of-plane vibration modes of C account for the majority of the contribution, while the in-plane vibration modes of C also make a minor contribution. At the medium-high frequency range of 580-1291 cm$^{-1}$, the in-plane vibration modes of C are predominant. The total and atom-projected PhDOS of C$_{2}$CaC$_{2}$ are depicted in Fig. 3(c). The Eliashberg spectral function $\alpha$$^{2}$$F$($\omega$) and the cumulative frequency-dependent EPC function $\lambda(\omega)$ are presented in Fig. 3(d). By comparing the total PhDOS in Fig. 3(c) with the Eliashberg spectral function $\alpha$$^{2}$$F$($\omega$) in Fig. 3(d), we can see that the peaks of them occur at similar frequency obviously. As shown in Fig. 3(d), the total EPC constant $\lambda$ of C$_{2}$CaC$_{2}$ is 0.75, half of which comes from the low frequency ($\omega$ \textless 352 cm$^{-1}$) part, with the remaining arising from mid-to-high frequencies.\\

\begin{figure*}
	\centering
	\includegraphics[width=1.0\linewidth,height=0.5\linewidth]{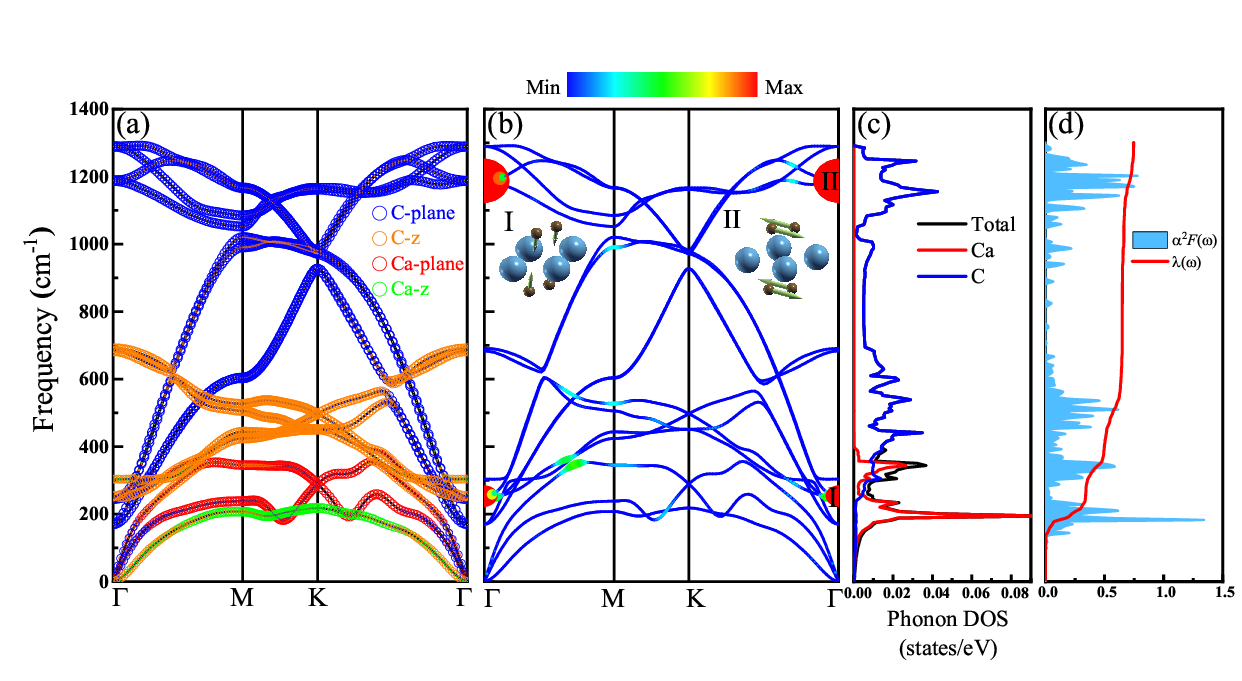}
	\caption{ (a) Phonon dispersion of C$_{2}$CaC$_{2}$ weighted by the vibrational modes of Ca and C atoms, respectively. (b) Phonon dispersion weighted by the magnitude of EPC $\lambda_{\mathbf{q}\nu}$. The insets show the vibration modes for the prominent $\lambda_{\mathbf{q}\nu}$ I and II, respectively. The brown and blue spheres represent C and Ca atoms, respectively. (c) Total and atom-projected phonon DOS for C$_{2}$CaC$_{2}$. (d) Eliashberg spectral function $\alpha$$^{2}$$F$($\omega$) and cumulative frequency dependence EPC function $\lambda(\omega)$ for C$_{2}$CaC$_{2}$. }
\end{figure*}
\begin{figure*}
	\centering
	\includegraphics[width=0.7\linewidth,height=0.65\linewidth]{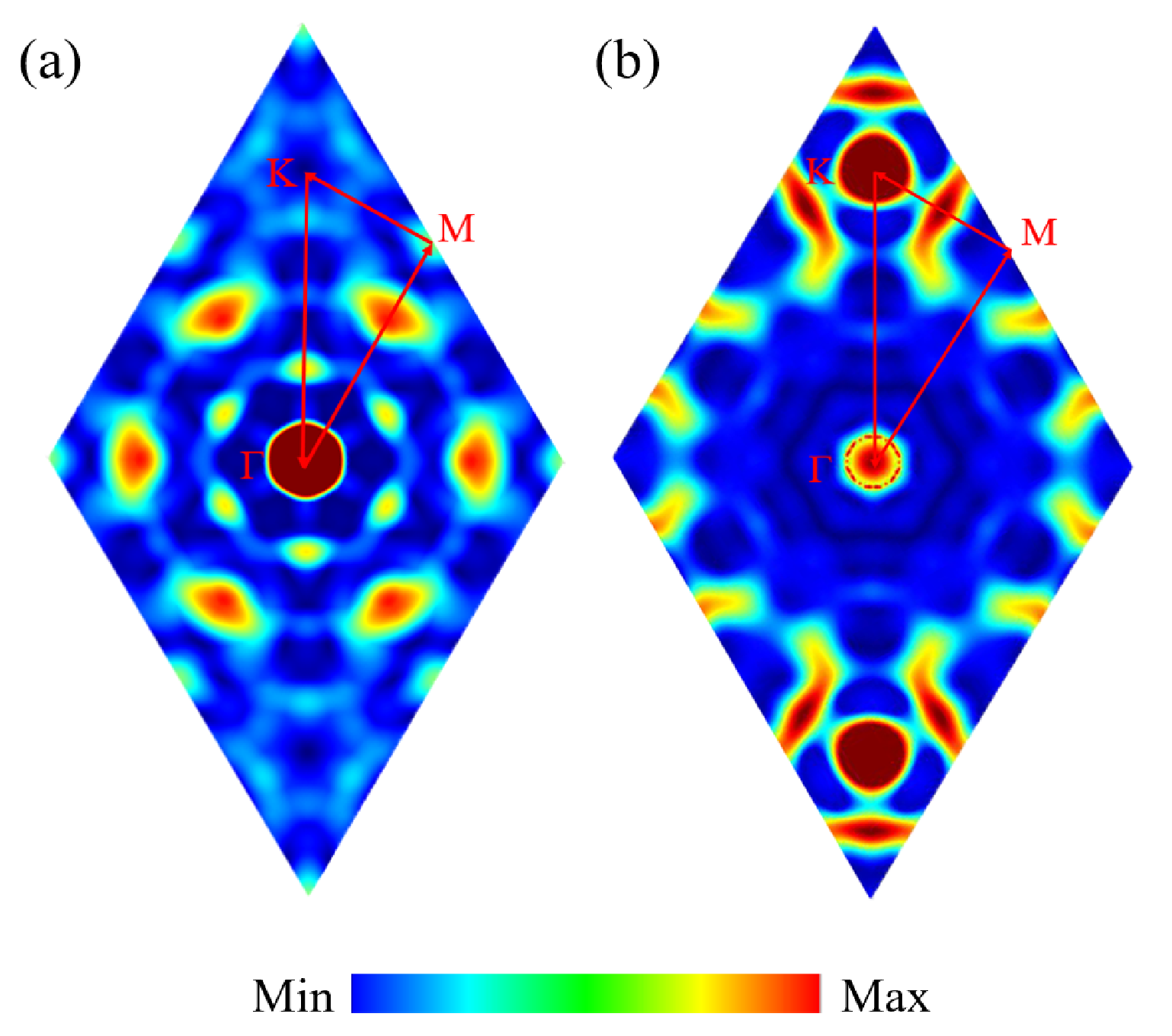}
	\caption{ The integrated EPC distributions of (a) pristine C$_{2}$CaC$_{2}$ in the plane of $q_{z}$ = 0. (b) The corresponding results for biaxial compressive strained ($\varepsilon$ = -4\%) C$_{2}$CaC$_{2}$. The high-symmetry path of $\Gamma$–$M$–$K$–$\Gamma$ is marked.}
\end{figure*}
\indent More specifically, for C$_{2}$CaC$_{2}$, as depicted in Fig. 3(a) and 3(d), approximately half of the EPC arises from the coupling of electrons and the low-frequency vibrations of Ca atoms below 352 cm$^{-1}$. And, the remaining EPC constant $\lambda$ mainly originates from the coupling between electrons, which mainly originate from the C-$p_{z}$ orbitals [Fig. 3(d)], and the in-plane and out-of-plane modes of C atoms ranging from 352 to 1291 cm$^{-1}$. Moreover, based on the magnitude of $\lambda_{\mathbf{q}\nu}$ shown in Fig. 3(b), it is evident that the strong coupling occurs in the vicinity of the $\Gamma$ point of the Brillouin zone, which corresponds well to Fig. 4(a). The strong coupling close to the frequencies of 255 cm$^{-1}$ (mode I) and 1202 cm$^{-1}$ (mode II), results from the coupling between electrons and the out-of-plane and in-plane vibrations of C atoms. The insets depicted in Fig. 3(b) illustrate the vibration for the two modes, with mode I corresponds to the out-of-plane vibration of C atoms, and mode II primarily corresponds to the in-plane vibration of C atoms.\\
\indent In 2D superconductors, such as 2D aluminum-deposited graphene AlC$_{8}$ \cite{ref:24} and the B-N compound B$_{3}$N \cite{ref:36}, applying biaxial strain typically results in an increase in the $T_{c}$. Thus, we investigate the potential impact of strain on the superconductivity of C$_{2}$CaC$_{2}$. The strain is applied along the two basis vector directions, and the relative change of the lattice constant $\varepsilon$ is defined as ($a$ $-$ $a_{0}$)/$a_{0}$ $\times$ 100\%, here $a_{0}$ and $a$ represent the lattice constants of pristine and strained C$_{2}$CaC$_{2}$, respectively. Positive/negative values represent tensile/compressive strains.\\

\begin{figure*}
	\centering
	\includegraphics[width=0.65\linewidth,height=0.5\linewidth]{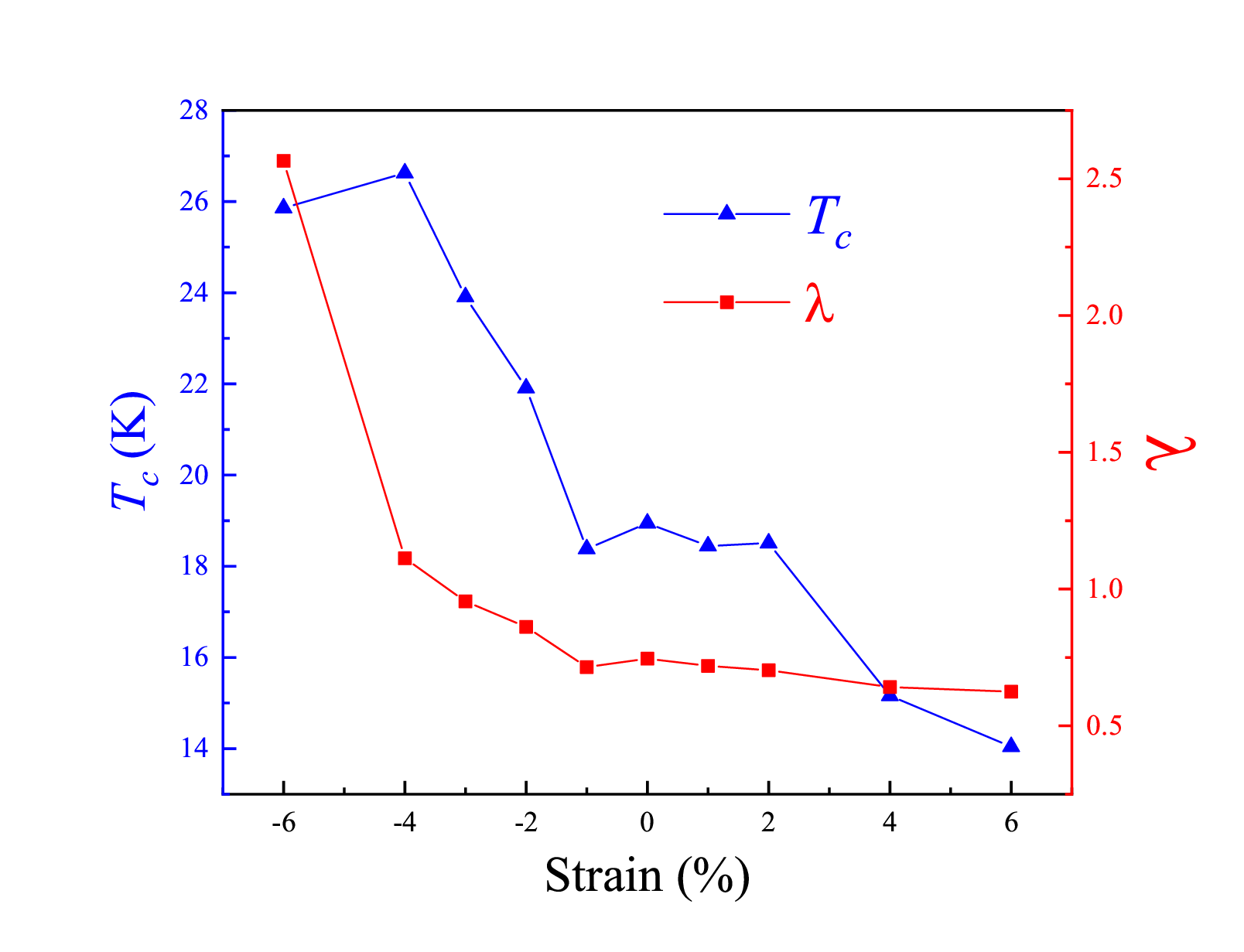}
	\caption{ Variations of $\lambda$ (red) and $T_{c}$ (blue) along with different strained cases.}
\end{figure*}
\begin{figure*}
	\centering
	\includegraphics[width=1.0\linewidth,height=0.5\linewidth]{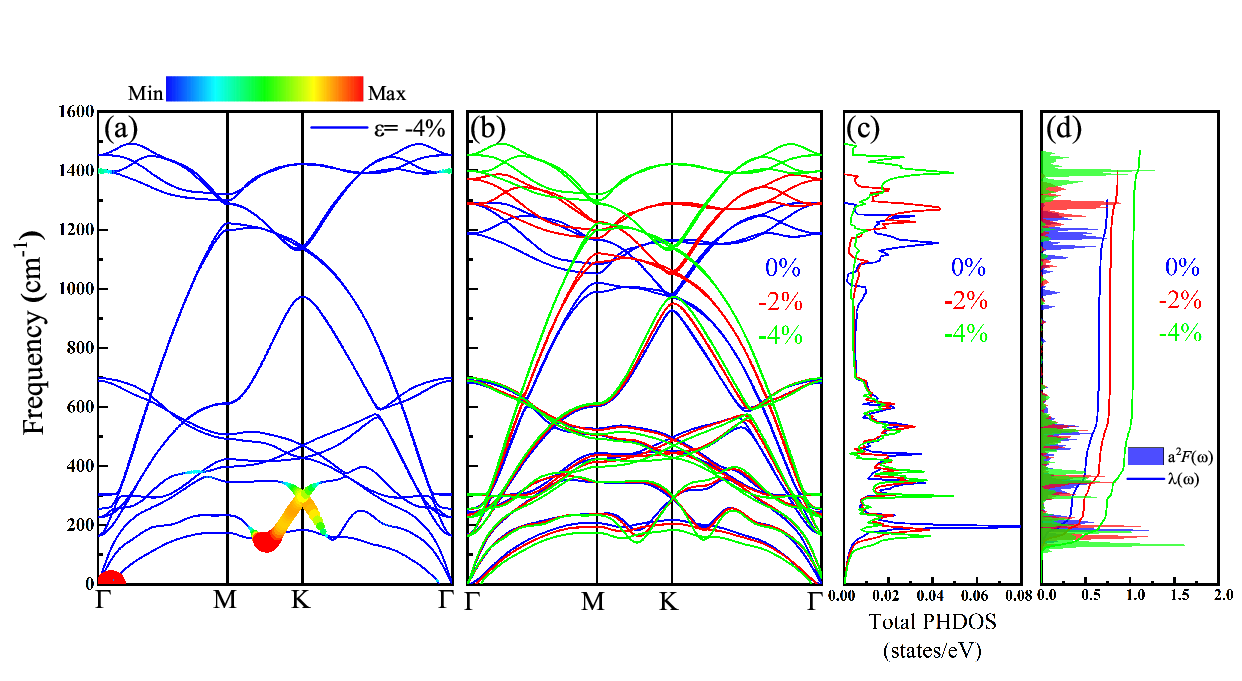}
	\caption{ (a) Phonon dispersion weighted by the magnitude of EPC $\lambda_{\mathbf{q}\nu}$ for C$_{2}$CaC$_{2}$ at compressive strain $\varepsilon$ = -4\%. (b), (c), and (d) are the Phonon dispersion, the total PhDOS, Eliashberg spectral function $\alpha$$^{2}$$F$($\omega$), and EPC function $\lambda(\omega)$ for the pristine, $\varepsilon$ = -2\%, and $\varepsilon$ = -4\% compressive strained cases, respectively.}
\end{figure*}
\begin{table*}
	\renewcommand\arraystretch{1.2}
	\tabcolsep=0.10cm
	\centering
	\caption{ List of superconducting parameters required for the calculation of $T_{c}$ in some reported 2D phonon-mediated graphene-based superconductors. This table includes data of doping, biaxial strain $\varepsilon$, logarithmic averaged phonon frequency $\omega_{log}$ (K), total EPC constant $\lambda$, and superconducting $T_{c}$ (K). Experimental $T_{c}$ values are also listed.}
	~\\ 
	\scalebox{0.65}{
		\begin{tabular}{cccccccccc}
			\hline 
			\hline 
			graphene-based materials & Doping & $\varepsilon$(\%) & $\omega_{log}$ (K) & $\lambda$ & $T_{c}$ (K) & Refs.\\ 
			\hline 
			AlC$_{8}$(mono)   & 0.2 hole per unit cell  & 12   & 188.452  & 1.557   & 22.23 & [25]\\ 
			
			LiC$_{6}$ (mono)     & 0.0  & 0.0 & 399.48    & 0.61      & 8.1 &[22] \\ 
			
			LiC$_{6}$ (mono)     & 0.0  & 0.0 & -    & 0.58$\pm$0.05      &   5.9 (Exp.) &[24] \\ 
			
			LiC$_{6}$ (bulk)     & 0.0  & 0.0    & 715.7      &0.33  & 0.9 &[22] \\ 
			
			CaC$_{6}$ (mono)     &0.0   & 0.0     & 445.64   & 0.4   & 1.4  &[22] \\ 
			
			CaC$_{6}$ (bulk)     &0.0   & 0.0     & 284.3   & 0.68   & 11.5 (Exp.)  &[7] \\ 
			
			C$_{6}$CaC$_{6}$     &0.0  &0.0 &- &-  & 2$\sim$4 (Exp.)      & [27] \\ 
			
			C$_{8}$KC$_{8}$   & 0.0 & 0.0  & -  & 0.46 & 3.6$\pm$0.1 (Exp.)   & [21] \\ 
			
			magic-angle graphene    &  0.0     & twist angle of 1.1°   & -  & -  & 1.7 (Exp.) &[2]  \\ 
			
			YbC$_{6}$(bulk)          & 0.0  & 0.0   & - & -   & 6.5 (Exp.) &[8] \\ 
			
			KC$_{8}$(bulk)      & 0.0 & 0.0    &-      &-     & 0.55 (Exp.)     & [26] \\ 
			
			C$_{2}$CaC$_{2}$    & 0.0 & 0.0    & 470.6   &  0.75     & 18.9       & This work \\ 
			
			C$_{2}$CaC$_{2}$    & 0.0 & -2    & 404.5   &  0.86     & 21.9       & This work \\ 
			
			C$_{2}$CaC$_{2}$    & 0.0 & -4     & 327.8   & 1.11   & 26.6      & This work \\ 
			\hline 
			\hline 
		\end{tabular}
	}
\end{table*}
\indent By calculation, we find that C$_{2}$CaC$_{2}$ is only stable when the strain is applied within the range of -6\% to 10\%. For the case of $\varepsilon$ = -7\% and 11\%, the phonon spectra exhibited imaginary frequencies, as depicted in Fig. B.10. Therefore, we only discuss superconductivity under stable strain conditions. When stretched, the $T_{c}$ of C$_{2}$CaC$_{2}$ decreases, as shown in Fig. 5. Conversely, when compressed, the $T_{c}$ increases. However, the $T_{c}$ no longer increases for $\varepsilon$ \textgreater -4\%. Fig. 6(a) illustrates the phonon dispersion weighted by the magnitude of EPC $\lambda_{\mathbf{q}\nu}$ at -4\% biaxial compressive strain, and Fig. 6(b-d) show the phonon dispersion, total PhDOS, Eliashberg spectral function $\alpha$$^{2}$$F$($\omega$), and EPC function $\lambda(\omega)$ for both pristine and -2\%, -4\% biaxial compressive strained cases. With an increase in compressive strain, low-frequency phonon softening and high-frequency phonon hardening of C$_{2}$CaC$_{2}$ can be obviously observed, as depicted in Fig. 6.(b). The phonon spectrum undergoes changes similar to those observed in other 2D materials under strain \cite{ref:37,ref:38}. In addition, it is worth mentioning that there are obvious Kohn anomalies in the path from $M$-$K$ and $K$-$\Gamma$ as depicted in Fig. 6.(a). These anomalies are commonly observed in other superconductors \cite{ref:39,ref:40} which significantly enhance $\lambda$ and ultimately increase $T_{c}$. When compressed by -2\%, $\lambda$ increases to 0.86, and the $T_{c}$ increases to 21.9 K. With a -4\% compression, $\lambda$ becomes 1.11, and the $T_{c}$ increases to 26.6 K, which is almost forty percent higher than the $T_{c}$ of the pristine C$_{2}$CaC$_{2}$ (18.9 K). The $T_{c}$ of the compressed C$_{2}$CaC$_{2}$ exceeds the liquid hydrogen temperature of 20.3 K. Besides, we plot the integrated EPC distributions for the case of -4\% biaxial compressive strain in Fig. 4(b), the strongest EPC region occurs around $K$, which is different from the pristine case in Fig. 4(a), a relatively large EPC region is also observed along $M$-$K$, which is consistent with the result in Fig. 6(a).\\

\section{Conclusions}
\indent In summary, based on first-principles calculations, we predicted a new calcium intercalated bilayer graphene C$_{2}$CaC$_{2}$, and studied its electronic structure, EPC, and possible superconductivity. The superconductivity of C$_{2}$CaC$_{2}$ mainly originates from the coupling between electrons from C-$p_{z}$ orbitals and the in-plane and out-of-plane vibration modes of the C atoms. The calculated EPC strength is approximately 0.75, and the superconducting $T_{c}$ is 18.9 K. The application of biaxial compressive strain, the $T_{c}$ can be boosted to 26.6 K, which is higher than other metal intercalated or deposited graphenes. It is anticipated that the predicted C$_{2}$CaC$_{2}$ and its strained cases can be realized in future experimental studies.\\
\section*{Acknowledgments} This work was supported by the National Natural Science Foundation of China (Grant Nos. 12074213, and 11574108), the Major Basic Program of Natural Science Foundation of Shandong Province (Grant No. ZR2021ZD01), the Natural Science Foundation of Shandong Province (Grant No. ZR2023MA082), and the Project of Introduction and Cultivation for Young Innovative Talents in Colleges and Universities of Shandong Province.\\

\appendix

\section {The crystal structure of calcium intercalated AB-stacking bilayer graphene C$_{2}$CaC$_{2}$}
\begin{figure*}
	\centering
	\includegraphics[width=1.0\linewidth,height=0.30\linewidth]{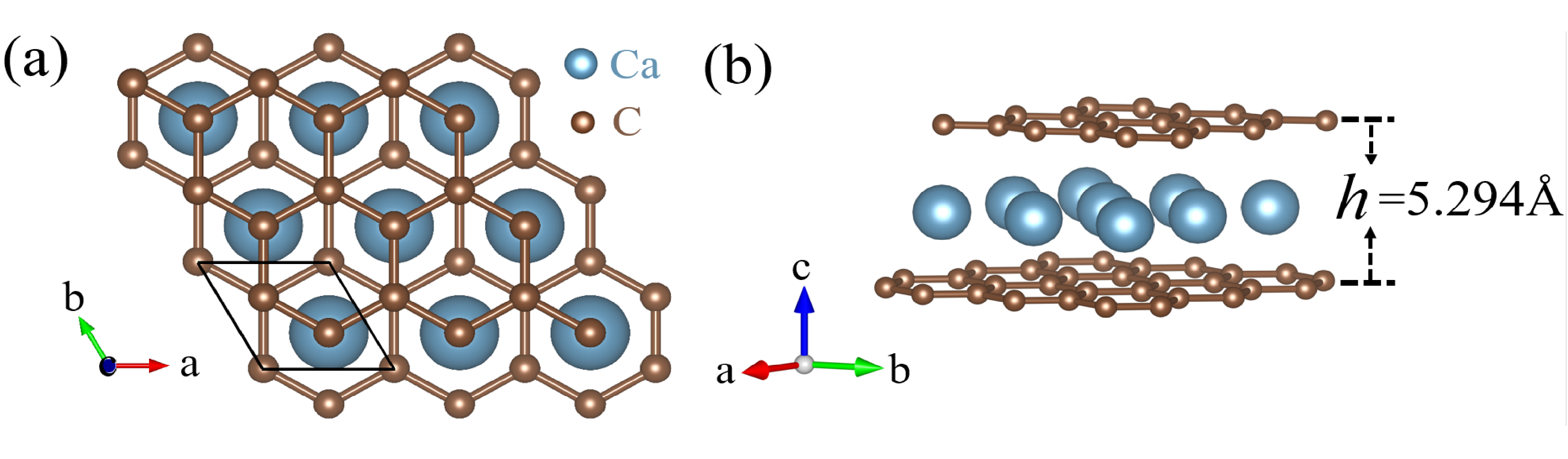}
	\caption{ Top (a) and side (b) views of calcium intercalated AB-stacking bilayer graphene C$_{2}$CaC$_{2}$. The unit cell is shown by the black solid line. Carbon and calcium atoms are represented by brown and blue spheres, respectively.}
\end{figure*}
\begin{figure*}
	\centering
	\includegraphics[width=1.0\linewidth,height=0.8\linewidth]{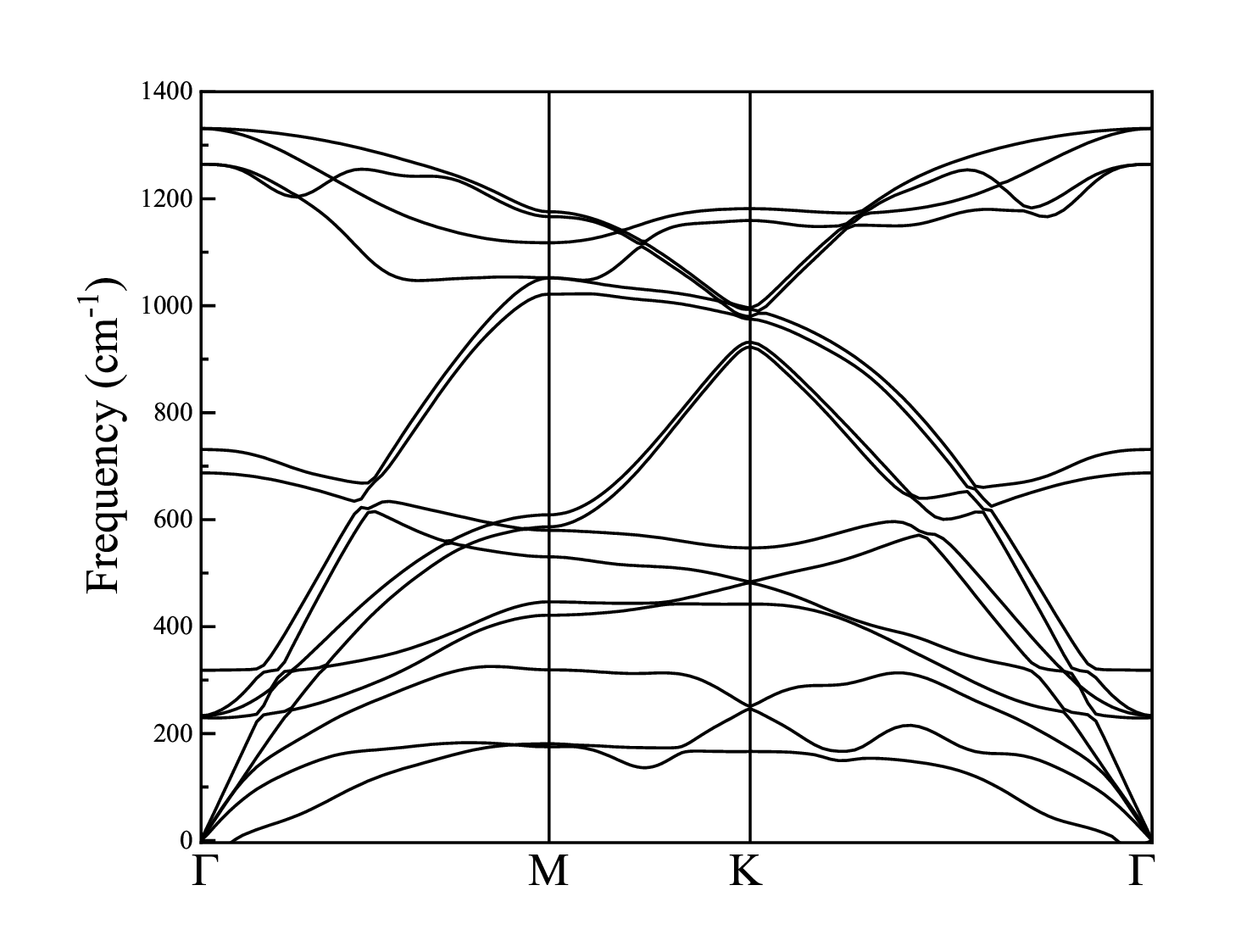}
	\caption{ Phonon dispersion for the AB-stacking C$_{2}$CaC$_{2}$.}
\end{figure*}
\begin{figure*}
	\centering
	\includegraphics[width=1.0\linewidth,height=0.75\linewidth]{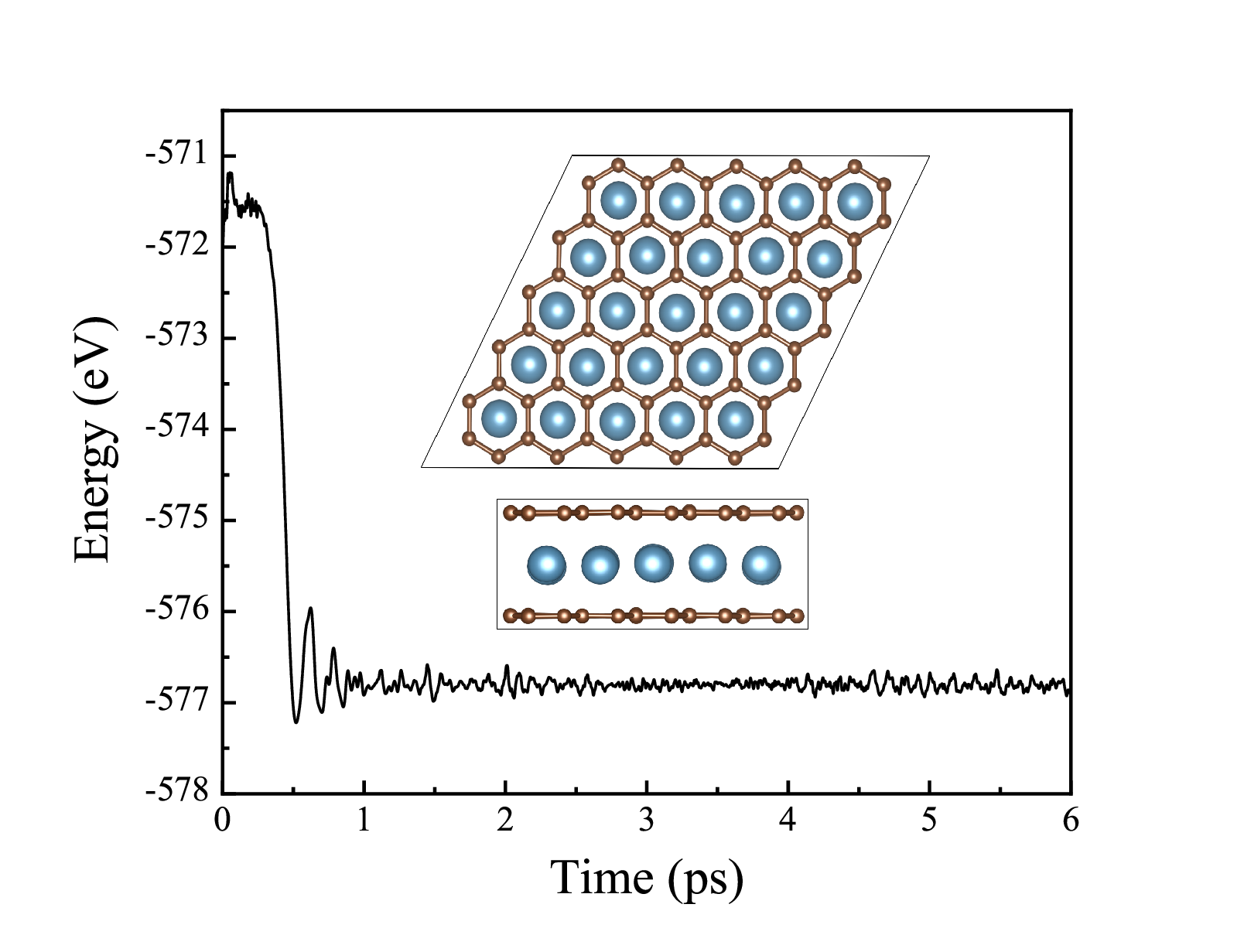}
	\caption{ The variation of the free energy in the AIMD simulations in the time scale of 6 ps along with the last frame of photographs at 50 K for AB-stacking C$_{2}$CaC$_{2}$.}
\end{figure*}
\indent The lattice structure of calcium intercalated AB-stacking bilayer graphene C$_{2}$CaC$_{2}$ is shown in Fig. A.7. After full optimization, the lattice parameter is 2.59 Å, which is slightly smaller than that of the AA-stacking C$_{2}$CaC$_{2}$. The height between the two graphene layers is 5.294 Å, slightly larger than that of the AA-stacking C$_{2}$CaC$_{2}$. When calculating its stability, we find that its phonon dispersion has no imaginary frequencies, as shown in Fig. A.8, indicating dynamic stability. However, during the subsequent verification of thermodynamic stability, it gradually changed to AA-stacking with lower system energy, as shown in Fig. A.9.\\

\section {The calculated phonon dispersion of C$_{2}$CaC$_{2}$ in the case of biaxial strain with $\varepsilon$ = -7\%, 11\%.}
\indent In the case of different strained cases applied to the structure, we find that with $\varepsilon$ = -7\% and 11\% of C$_{2}$CaC$_{2}$, the structure begins to show dynamic instability. As is shown in Fig. B.10, the phonon dispersion exhibits imaginary frequencies.\\
\begin{figure*}
	\centering
	\includegraphics[width=0.9\linewidth,height=0.7\linewidth]{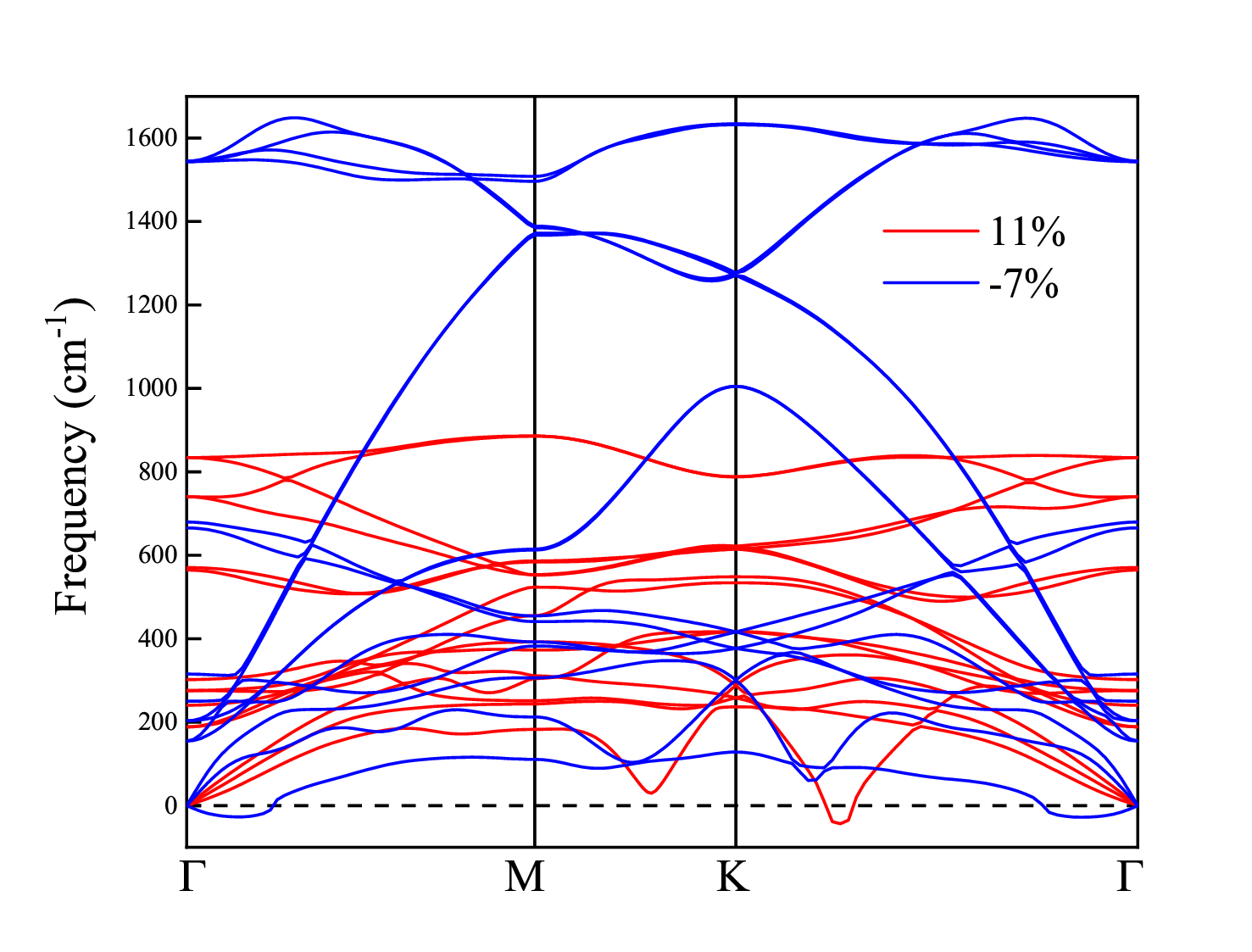}
	\caption{ Phonon dispersion of C$_{2}$CaC$_{2}$ in the case of biaxial strain with $\varepsilon$ = 11\% and -7\%.}
\end{figure*}

\section {The $T_{c}$ of pristine calcium intercalated AA-stacking bilayer graphene C$_{2}$CaC$_{2}$ and biaxial compressive strained C$_{2}$CaC$_{2}$ as a function of Coulomb pseudopotential $\mu^{*}$}
\begin{figure*}
	\centering
	\includegraphics[width=0.95\linewidth,height=0.7\linewidth]{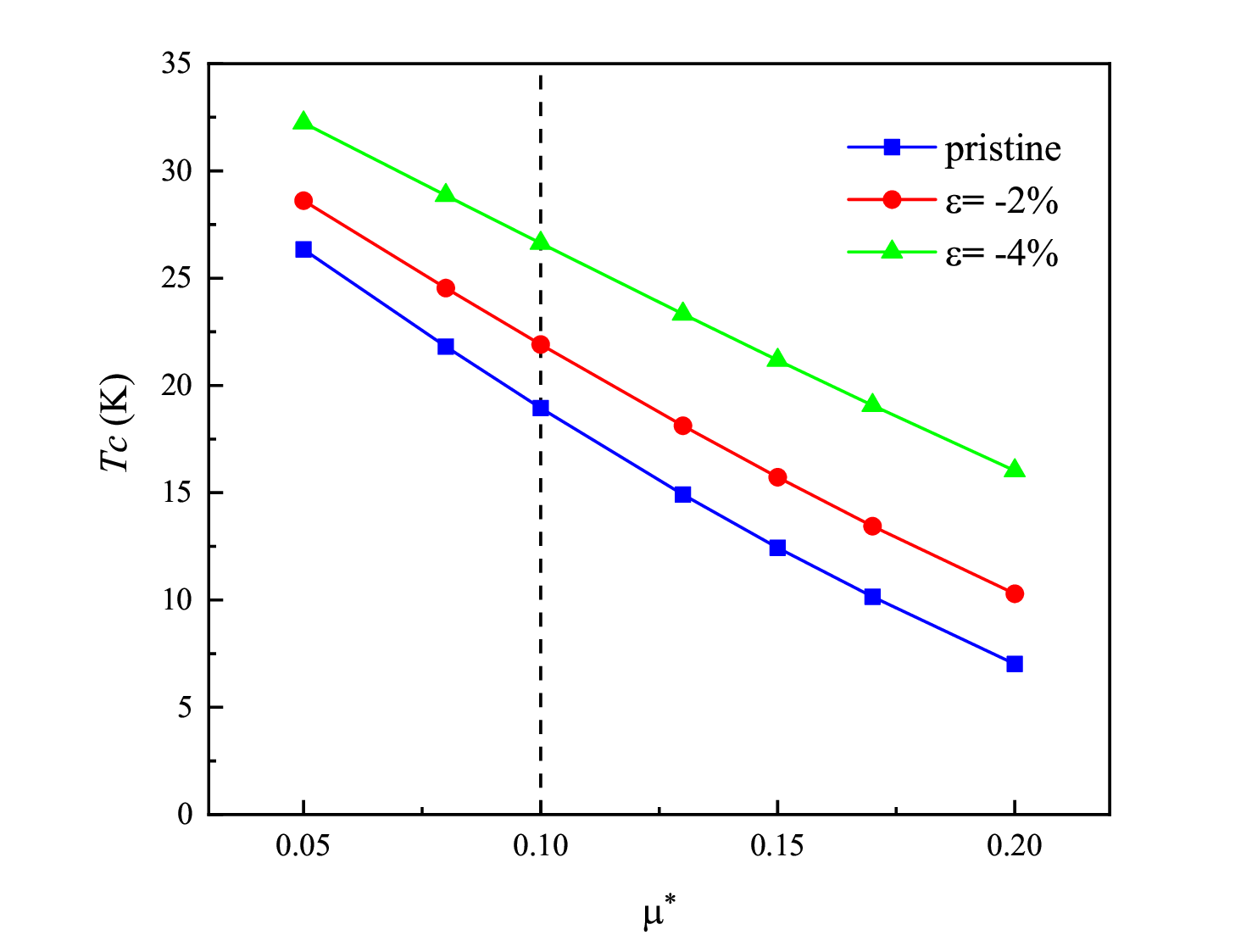}
	\caption{ The calculated $T_{c}$ of pristine calcium intercalated AA-stacking bilayer graphene C$_{2}$CaC$_{2}$ and biaxial compressive strained C$_{2}$CaC$_{2}$ as a function of Coulomb pseudopotential $\mu^{*}$. The vertical line marks the value of $\mu^{*}$ = 0.10 used in the main text.}
\end{figure*}
\indent It is well known that the Coulomb pseudopotential $\mu^{*}$ is an empirical parameter that is closely associated with $T_{c}$. The calculated $T_{c}$ of pristine calcium intercalated AA-stacking bilayer graphene C$_{2}$CaC$_{2}$ and biaxial strained C$_{2}$CaC$_{2}$ as a function of $\mu^{*}$ is shown in Fig. C.11. The value of $\mu^{*}$ is considered in the range of 0.05 $–$ 0.20. As is seen in Fig. C.11, the $T_{c}$ decreases monotonically with the increasing of $\mu^{*}$ for both pristine C$_{2}$CaC$_{2}$ and biaxial compressive strained C$_{2}$CaC$_{2}$. As $\mu^{*}$ increases from 0.05 to 0.20, $T_{c}$ decreases from 26.3 K, 28.6 K, 32.2 K to 7.1 K, 10.3 K, 16.1 K for the pristine, -2\%, and -4\% compressive strained cases, respectively. For the commonly used $\mu^{*}$ = 0.10, the $T_{c}$ for the pristine, -2\% and -4\% compressive strained cases are 18.9 K, 21.9 K, and 26.6 K, respectively, which are listed in Table 1 of the main text.\\

\end{document}